\documentclass[reprint, aps, pre]{revtex4-2} 
\usepackage[title]{appendix}
\usepackage{amsmath, amssymb, bm}
\usepackage{graphicx, dcolumn, enumitem, xcolor}
\graphicspath{{figures/}}
\usepackage{hyperref}
\usepackage{comment}
\usepackage{enumitem}
\usepackage{gensymb}

\allowdisplaybreaks

\let\tempvec\vec
\renewcommand{\vec}[1]{\tempvec{{}#1}}
\let\tempbar\bar

\renewcommand{\bar}[1]{\tempbar{{}#1}}
\newcommand{\pmat}[2]{\def\arraystretch{#2}\begin{pmatrix}#1\end{pmatrix}} 
\newcommand{\w}[1]{\omega^{#1}} 
\newcommand{\X}[1][]{\bm{x}^{3D}_{#1} \paren{\w{1}, \w{2}, \w{3}}} 
\newcommand{\xmid}[1][]{\bm{x}^{m}_{#1}(\w{1}, \w{2})} 
\newcommand{\normalVec}[1][]{\bm{\hat{N}}^m_{#1} \paren{\w{1}, \w{2}}} 
\newcommand{\paren}[1]{\left( #1 \right)}

\newcommand{\strain}[2]{\gamma_{#1#2}} 
\newcommand{\gbar}[2]{\bar{g}_{#1#2}} 
\newcommand{\gcbar}[2]{\bar{g}^{#1#2}} 
\newcommand{\gind}[2]{g_{#1#2}} 
\newcommand{\pder}[2]{\partial_{#1} \bm{#2}} 
\newcommand{\tangent}[1]{\partial_{#1}\bm{x}^{3D}} 
\newcommand{\hab}{h_{\alpha \beta}}

\begin{document}
	\title{Twist-driven helical flattening in nematic elastomer cylinders}
	\author{Alexia Chatzitheodorou}
		\email{achatzit@syr.edu}
	\author{Christian D. Santangelo}
        \email{cdsantan@syr.edu}
	\affiliation{Department of Physics, Syracuse University, Syracuse, New York 13244, USA}
	\date{\today}
	
	\begin{abstract}
        Liquid Crystal elastomers (LCEs) deform anisotropically along a prescribed nematic director field, making them promising candidates for programmable shape change. Existing studies have primarily focused on  either director patterns on flat sheets or simple patterns on curved geometries, leaving the case of a non-trivial director on a non-trivial geometry still largely unexplored. Many biological systems, however, have both complex geometries and complex, helical fiber architectures. To explore the interplay between director and underlying geometry, we use a non-Euclidean plate theory for nematic elastomers with a through-thickness twisted director to develop an effective 2D model. With a fixed twist angle, our model shows an anomalous coupling between mid-surface curvature and director twist. This emergent term arises from the interplay between orientational order and elasticity, and dominates traditional bending contributions. To illustrate the general theory, we study the stability of cylindrical shapes with through-thickness twist. We find that the cylinder is unstable to a long-wavelength helical flattening mode and determine the critical parameter for the onset of instabilities.
	\end{abstract}
	\maketitle

\section{Introduction}\label{sec:Introduction}
Biological systems undergo dramatic shape changes during development. Embryos fold during gastrulation \cite{morphogenesismodelsreview}, leaves and flowers curl as they grow \cite{longleafMaha}, seed pods snap open \cite{seedpodsSharon}, and organs deform under stress, from breathing lungs to the contracting heart and the peristaltic gut \cite{growthbioGoriely, growthbioLee}. These large deformations can be driven by internal stresses induced by differential growth 
that cannot be globally relaxed, a phenomenon known as geometric incompatibility \cite{growthbioGoriely, growthcontinuumMcculloch}. In that case, the resulting internal stresses are released through mechanical instabilities such as buckling \cite{gutbucklingMaha}, snap-through \cite{venusflytrapForterre}, and wrinkling \cite{wrinklingMaha}. 

Inspired by these biological mechanisms, a class of synthetic shape-morphing materials has emerged in which geometric incompatibility is programmed at fabrication in order to trigger a complex, global shape change. Among the systems studied are hydrogels encoded with patterns of differential swelling \cite{gelsKleinSharon, gelsRyanChris, gelsSabettaMaha}, baromorphs containing patterns of inflatable channels \cite{pneumaticfabricsSiefert, pneumaticelastomersSiefert}, liquid crystal elastomers \cite{LCEsWarner}, self-folding origami and kirigami structures \cite{origamiNaChris}, and many more. These systems share a common design principle: local deformations are patterned on an initially flat sheet and the subsequent three dimensional shape that emerges is minimizing the material's elastic energy.

Liquid Crystal Elastomers (LCEs), in particular, are a heavily studied class of shape-morphing materials. They are cross-linked polymer networks of rod-like mesogens that
align along a common orientational axis, forming a nematic phase. The axis is captured by the nematic director, $\hat{\bm n}$, which possesses head-to-tail symmetry ($\hat{\bm n}\equiv-\hat{\bm n}$)  \cite{LCEsWarner}.
When external stimuli, such as heat or light, disrupt the nematic order, 
the network contracts along the director $\hat{\bm n}$ and elongates transverse to it.
Since the director governs this response locally, a spatially-varying $\hat{\bm n}$ results in a spatially varying deformation, setting the ultimate geometry after actuation. Therefore, there is a direct association between heterogeneous cross-link density and the buckled geometry.

The relationship between the prescribed director field and the resultant geometry has been widely studied: uniform director fields produce simple uniaxial, flat deformation \cite{uniformKupfer}, while planar director fields $\hat{\bm n}(x,y)$ that are uniform through the sheet's thickness generate Gaussian curvature, driving flat sheets into non-trivial geometries when actuated \cite{saddlesconesWhite, AharoniNematic}. Somewhat less explored are director textures that vary through the thickness, but such through-thickness gradients are widespread in biological systems with helically arranged fiber architectures, such as the heart muscle \cite{ZhangHeart}, the Bouligand structure of the mantis shrimp dactyl \cite{SuksangpanyaMantisShrimp}, and the twisted plywood arrangement of collagen fibers in human cortical bones \cite{ReisingerBone}.
Most theoretical and experimental work in this regime, however, has focused on initially two-dimensional LCE sheets \cite{twistedSelinger}; far less is known about how a twisted director couples to a reference geometry that is non-trivial,  the regime in which biological systems most naturally sit. 

In this paper, we take up this question for a cylindrical nematic elastomer with a director that twists through its thickness, a geometry loosely inspired by the helical fiber organization of the heart muscle \cite{MahaHeartpump}. We will show that, because the cylindrical topology frustrates the ability of the sheet to achieve its locally preferred twisted geometry, the resulting deformation drives a change in the cylinder's equilibrium radius. We show that this frustration renders the equilibrium state unstable to a long-wavelength helical flattening mode: the cross-section becomes elliptical and it rotates along the cylinder axis, such that the through-thickness twist of the director appears as a twist in the modulated cross-section itself.

Classical plate theories – including F\"oppl-von K\`arm\`an \cite{elasticityLandau, elasticityAudoly} and Koiter \cite{Koiter} – measure strain as a deviation from a flat, stress-free configuration. This assumption fails in the context of differential growth when there is no stress-free reference state. A more general framework is provided by non-Euclidean (or incompatible) elasticity \cite{efratiElasticTheoryUnconstrained2009}, a framework that has been used to study growth in both hydrogels \cite{gelsKleinSharon} and nematic elastomers. This approach has been facilitated by dimensional reduced theories with rigorous footing through $\Gamma$-convergence \cite{gammaPlucinsky, gammaAgostiniani} as well as metric formulations.

In this paper, we construct a non-Euclidean theory for a thin nematic elastomer plate with a director that twists rapidly through its thickness. 
To our knowledge, no such reduction has been carried out for a twisted through-thickness director on a non-trivial reference geometry. Starting from a three-dimensional energy functional, we perform a dimensional reduction in curvilinear coordinates. In addition to the standard stretching (order $t$) and bending (order $t^3$) contributions of Kirchhoff-Love theory, the reduction yields an anomalous $t^2$ energy contribution. 
We then apply the resulting theory to the buckling of a cylinder and find that the cylinder becomes unstable.
The paper is organized as follows: in Sec. \ref{sec:EffectiveTheory} we derive the effective theory and  set up the perturbation analysis, in Sec. \ref{sec:Results} we present the results of a linear stability analysis on a cylindrical geometry. We conclude in Sec. \ref{sec:Conclusion} with a discussion of future directions in the shape-programming of twisted nematic elastomers.

\section{Effective Elasticity Theory for Twisted Nematic LCEs}\label{sec:EffectiveTheory} 

\subsection{Geometric description}\label{subsec:Geom}

We model the liquid crystal elastomer (LCE) plate in terms of a local director field, $\hat{\bm n}$, which is tangent to the inner and outer surfaces of a thin shell of thickness $t$. We assume that the material is cross-linked deep in the nematic phase, and adopt the strong volumetric anchoring limit in which the director follows the deformations of the network instead of being an independent dynamical field. We also neglect Frank elasticity, which penalizes spatial variations in the director field, as its effects are expected to be small compared to the elastic energy \cite{Mihai2021}.
Elastic bodies that have no stress-free configuration can be described using the framework of non-Euclidean plates, also known as incompatible elasticity \cite{efratiElasticTheoryUnconstrained2009}. The framework assumes that the mechanical state of such a system is characterized by two distinct metrics. The induced, or actual metric, $\gind{i}{j}$, describes the geometry actually realized by the deformed configuration. The reference, or target metric, $\gbar{i}{j}$, on the other hand, is intrinsic to the material and encodes the local distances that would render the system locally stress-free.
The strain tensor is defined as 
\begin{equation}
    \strain{i}{j}=\frac{1}{2}\paren{\gind{i}{j}-\gbar{i}{j}}.
\end{equation}
We adopt the Einstein summation convention whereby repeated indices are summed. We use Latin indices \textit{i,j,k,..} $\in \{1,2,3\}$ for tensors defined on the three dimensional body and Greek indices \textit{$\alpha, \beta, \gamma,..$}$\in \{1,2\}$ for tensors defined on the mid-surface. Indices throughout this work are raised and lowered with $\gbar{i}{j}$, which serves as the natural geometry for measuring elastic deformations.
We assume the reference metric takes the block-diagonal form
\begin{equation}
    \gbar{i}{j} \;=\; \pmat{\gbar{\alpha}{\beta}^{3D} & 0 \\ 0 & t^2}{1}.
\end{equation}
In index notation, this means that $\gbar{\alpha}{3}=0$, $\gbar{3}{3}=t^2$.

\begin{figure}[t]
    \centering
\includegraphics{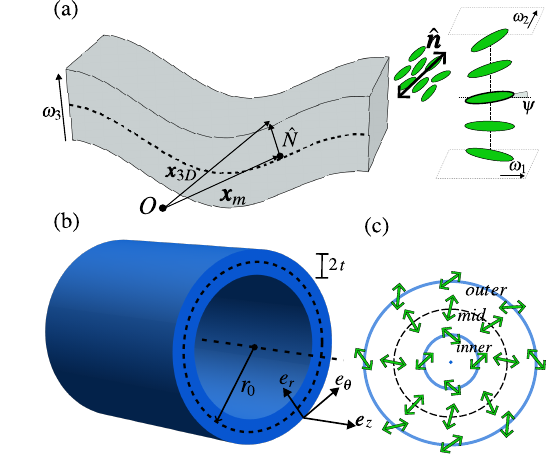}
    \caption{Geometry of a twisted nematic elastomer plate. (a) Three-dimensional view of a curved plate with the midsurface shown (dashed). The director (green ellipsoids) rotates through the thickness. $\psi$ is the midsurface pitch angle between the director at $\w{3}=0$ and the in-plane basis vector along $\w{1}$. Here we illustrate $\psi=10^{\circ}$ at the midsurfarce and a total rotation $2\pi \beta$ across the thickness set to $60^{\circ}$ from inner to outer surface. (b) Cylindrical reference geometry with midsurface of radius $r_0$ and total thickness $2t$ with midsurface dashed. The physical coordinates $(r,\theta,z)$ correspond to curvilinear coordinates $\w{1},\w{2},\w{3}$ of general framework. (c) Cross view in the $\bm{ e}_r,\bm{e}_{\theta}$ plane, showing the $60^{\circ}$ through-thickness rotation as in panel (a).}

    \label{fig:cylGeom}
\end{figure}

The three-dimensional embedding $\X$ of the deformed plate is parametrized by curvilinear coordinates $\w{1},\w{2},\w{3}$ with $\w{1}, \w{2}$ the in-plane coordinates, and $\w{3}\in[-1,1]$ a dimensionless coordinate along the thickness, as seen in Fig. \ref{fig:cylGeom}. Assuming the Kirchhoff-Love kinematic constraints, which require no transverse shear and no normal strain, $\gamma_{i3}=0$,
the embedding is,
\begin{equation} \label{eq:3Dembedding}
\begin{aligned}
    \X
    &=\xmid \\
    &+ \w{3} t \normalVec,
\end{aligned}
\end{equation}
where $\xmid$ is the midsurface immersion and $\normalVec[]$  is the midsurface normal vector.
The induced surface is defined by $\gind{i}{j}=\tangent{i}\cdot\tangent{j}$ and represents an unknown quantity that the deformed plate must achieve. The induced mid-surface carries a second fundamental form  $\hab=-~\pder{\alpha}{x}^{m}\cdot \pder{\beta}{\bm{\hat{N}}}^{m}$ that encodes the local curvature.


In tandem with the Kirchoff-Love approximation, we also assume the director remains tangent to the midsurface of the thin plate, i.e. $n^3=0$.
The coordinate basis $\{\partial_1,\partial_2,\partial_3\}$ induced by the embedding $\X$ is in general not orthonormal. Therefore, to define the director field $\bm{n}$, we introduce an orthonormal frame $\{\bm{e}_1,\bm{e}_2,\bm{e}_3 \}$ on the materials' coordinates where $\bm{e}_3 = \normalVec[]$ is pointing along the normal direction, $\bm{e}_1$ is the projection of an in-plane coordinate tangent basis vector $\partial_1$ onto the tangent plane normalized to unit length, and $\bm{e}_2~=~\bm{e}_3\times\bm{e}_1$ completes the frame. Each in-plane frame vector can be written as $\bm{e}_i =e_i^\alpha \partial_\alpha$ where $e_1^{\alpha},~ e_2^{\alpha}$ are the components of the orthonormal frame expressed in the coordinate basis, with the lower index indicating the coordinate index of the orthonormal frame vector and the upper index indicating the component of that frame vector.
The three-dimensional planar unit vector director field defined in the reference state can be parametrized as $\hat{\bm{n}}\paren{\w{1},\w{2},\w{3}}=\cos\paren{\Phi}\bm{e}_1+\sin\paren{\Phi}\bm{e}_2$.
We use $\Phi=\pi\beta\w{3}~+~\psi$ to model uniform twisting of the director from the inside to the outside, leading to contravariant components of the form,
\begin{align}\label{eq:3Dnematic}
    n^{\alpha}  &=~\cos\paren{\pi\,\beta\,\w{3}~+~\psi}e_{1}^{\alpha}\\ \nonumber
                &~+~\sin\paren{\pi\,\beta\,\w{3}~+~\psi}e_{2}^{\alpha}.
\end{align}
The twist angle, $\pi \beta$, is controlled by a dimensionless parameter, $\beta$. 
The mid-surface pitch angle $\psi$ specifies the orientation of the director at $\w{3}=0$ relative to $\bm{e}_1$, as seen in Fig. \ref{fig:cylGeom}. 

The liquid crystal tensor which characterizes the nematic order, is defined by
\begin{equation}\label{eq:Qtensor}
    Q^{ij}=q \left( n^i n^j - \frac{1}{3} g^{i j} \right),
\end{equation}
where $q$ is the scalar order parameter that shows the degree of orientation of the LC molecules, $n^i$ is the nematic director normalized such that $g_{ij}n^i n^j=1$. By construction, $Q^{ij}$ is symmetric and traceless with respect to the induced metric, $g_{ij} Q^{ij}=0$. This definition reflects a deliberate choice: the trace condition is imposed using the actual metric $g$ rather than the reference metric because the nematic director describes the orientational distribution of mesogens in the current state without a memory of the reference configuration. This is consistent with the Eulerian perspective in which material properties are expressed in terms of the current state. In fact, this choice and the alternative $\bar{g}_{ij} Q^{ij}=0$ produce the same terms in the energy up to leading order.

\subsection{Dimensional reduction}\label{subsec:DimRed}
For thin bodies, the three-dimensional elastic energy can be recast into an effective energy of a mid-surface alone. We carry out this reduction within the non-Euclidean framework introduced in Sec.~\ref{subsec:Geom} and using the Kirchhoff-Love approximation.
Our starting point is a three-dimensional energy functional consisting of two contributions, an elastic energy based on isotropic Hookean elasticity, $\mathcal{E}_{elastic}^{3D}$, and a nematic-elasticity coupling term, $\mathcal{E}_{\gamma\,-\,Q}^{3D}$, that captures the interaction between strain measured relative to a reference state, and the instantaneous orientational order \cite{LCEsWarner, calderer2013landaugennestheoryliquid}. The coupling favors strain aligned with the director. As discussed, we do not include any terms associated with the Frank energy of a nematic liquid crystal since these effects are expected to be small compared to the elastic energy \cite{Mihai2021}. The energy density thus take the form
\begin{equation} \label{eq:3DEnergy}
    \mathcal{E}^{3D}=\frac{1}{2}A^{ijkl}\strain{i}{j}\strain{k}{l} ~+~ \eta\,\strain{i}{j}Q^{ij},
\end{equation} 
where $A^{ijkl}=\lambda~\gcbar{i}{j}\gcbar{k}{l}~+~\mu~\paren{\gcbar{i}{k}\gcbar{j}{l}~+~\gcbar{i}{l}\gcbar{j}{k}}$ is the elastic stiffness tensor that depends only on material properties \cite{elasticityLandau} with $\lambda,~\mu >0$ the Lam\'e coefficients. $\eta$ controls the coupling strength with units of energy density and same order of magnitude as the shear modulus $\mu$ \cite{deGennes1975}.

From the embedding in Eq. \ref{eq:3Dembedding}, the components of the 
induced three-dimensional metric, $\gind{\alpha}{\beta}^{3D}$, $\gind{\alpha}{3}^{3D}\;$ and $\gind{3}{3}^{3D}$ are computed as power expansions in the thickness coordinate $\omega^{3}$
\begin{eqnarray}
    \gind{\alpha}{\beta}^{3D} &=& \partial_{\alpha}\bm{x}^{3D}\,\cdot\,\partial_{\beta}\bm{x}^{3D} \nonumber \\
    &=& \gind{\alpha}{\beta} \;-\; 2(\w{3}t)\hab \;+\; (\w{3}t)^{2} g^{\lambda \kappa} h_{\alpha \lambda} h_{\beta \kappa}, \\
    \gind{\alpha}{3}^{3D}
        &=&\pder{\alpha}{x}^{3D} \cdot \pder{3}{x}^{3D} = 0, \nonumber \\
    \gind{3}{3}^{3D} &=& 1,\nonumber
\end{eqnarray}
where thickness is set to a constant and $\hab=-~\pder{\alpha}{x}^{m}\cdot \pder{\beta}{\bm{\hat{N}}}^{m}$ is the mid-surface second fundamental form. Following the earlier assumption $\bar{h}_{\alpha \beta}=0$, we have  $\bar{g}^{3D}_{\alpha \beta}\approx\bar{g}_{\alpha \beta}$ and the strain therefore decomposes to
\begin{equation} \label{eq:strainExp}
    \strain{\alpha}{\beta}^{3D} = \frac{g_{\alpha \beta}-\bar{g}_{\alpha \beta}}{2} - \w{3} t\, h_{\alpha \beta} + (\w{3}t)^2 \frac{g^{\lambda \kappa} h_{\alpha \lambda} h_{\beta \kappa}}{2},
\end{equation}
where $\gamma^{(0)}$ is the mid-surface strain, $\gamma^{(1)}$ encodes bending through the mid-surface curvature, and $\gamma^{(2)}$ is the third fundamental form.

We perform the dimensional reduction in two steps. Starting with the elastic energy, we apply the plane-stress assumption $\sigma_{i3}=0$, which yields a condition for the normal strain $\gamma_{33}=-[\lambda t^2/(\lambda+2\mu)]\gamma^\alpha_\alpha$, where we assume that $q$ is relatively small. Assuming no transverse shear $\strain{\alpha 3}~=~0$, the elastic energy density can be written in terms of in-plane strains only, $\mathcal{E}_{elastic}^{3D}=~\mu \gamma_{\alpha\beta}\gamma^{\alpha\beta} + \mu \frac{\lambda}{\lambda+2\mu}\paren{\gamma^{\alpha}_{\alpha}}^2$. Retaining terms up to the second order in  $\w{3}$, and integrating through the thickness, we obtain both stretching and bending contributions to the elastic energy consistent with classical Kirchhoff-Love plate theory,
\begin{equation}
    \mathcal{E}_{elastic}=t\mathcal{E}_{s}+t^{3}\mathcal{E}_{b}
\end{equation}
where 
\begin{align} \label{eq:elasticReduced}
\mathcal{E}_s~
&=~\mu \paren{\gamma^{\alpha\beta}_{(0)}\strain{\alpha}{\beta}^{(0)}~+~\frac{\lambda}{\lambda+2\mu}\paren{\gcbar{\alpha}{\beta}_{(0)}\strain{\alpha}{\beta}^{(0)}}^2} \nonumber \\
\mathcal{E}_b~
&=~\mu \left( \underbrace{\gamma^{\alpha\beta}_{(0)}\strain{\alpha}{\beta}^{(2)}}~+~\gamma^{\alpha\beta}_{(1)}\strain{\alpha}{\beta}^{(1)}~+~\frac{\lambda}{\lambda+2\mu} \paren{\gcbar{\alpha}{\beta}_{(0)}\strain{\alpha}{\beta}^{(1)}}^2  \right. \nonumber \\
&\left. +\underbrace{~\gcbar{\alpha}{\beta}_{(0)}\strain{\alpha}{\beta}^{(0)}\gcbar{\mu}{\nu}_{(0)}\strain{\mu}{\nu}^{(2)}} \right).
\end{align} 
Assuming small deviations, it is reasonable to drop the first and last terms of the bending energy that appear in Eq. \ref{eq:elasticReduced}.

Following an analogous procedure, we can obtain the effective two-dimensional strain-liquid crystal energy $\mathcal{E}_{\gamma-Q}^{3D}$. We focus only on contributions from the strain-nematic term $\mathcal{E}_{\gamma\,-\,n}=\gamma_{\alpha \beta}n^{\alpha}n^{\beta}$. 
The nematic tensor product $n^{\alpha}n^{\beta}$ that appears in Eq. \ref{eq:Qtensor} can be written in the orthonormal frame as 
\begin{equation}\label{eq:nematicPrd}
    n^{\alpha}n^{\beta}~=~\frac{1}{2}g^{\alpha \beta}_{3D}~+~\frac{1}{2}\cos{2\Phi}\Delta^{\alpha \beta}~+~\frac{1}{2}\sin{2\Phi}S^{\alpha \beta},
\end{equation}
where $g^{\alpha \beta}_{3D}=e_1^\alpha e_1^\beta+e_2^\alpha e_2^\beta$, $\Delta^{\alpha \beta}=e_1^\alpha e_1^\beta-e_2^\alpha e_2^\beta$, and $S^{\alpha \beta}=e_1^\alpha e_2^\beta+e_2^\alpha e_1^\beta$. The tensor product depends on the thickness coordinate through both the twist angle $\Phi=\pi\beta\w{3}+\psi$ and the orthonormal frame basis vector components $e_{i}^{\alpha}$. The frame basis vectors can be expanded as polynomials $e_i^\alpha e_j^\beta=e_{i(0)}^\alpha e_{j(0)}^\beta +\w{3}t\paren{e_{i(0)}^\alpha e_{j(1)}^\beta+e_{i(1)}^\alpha e_{j(0)}^\beta}$, where $e_{i(0)}^\alpha$ are the mid-surface frame components, and $e_{i(1)}^\alpha$ the corrections. In contrast, the trigonometric factors are not Taylor-expanded because we assume a rapidly varying director. This will be the source of an anomalous $t^2$ contribution we identify later.

Combining the frame expansion with the strain expansion in Eq.~\ref{eq:strainExp}, we retain up to quadratic order in $\omega^{3}$ of the strain-nematic term $\gamma_{\alpha \beta}n^{\alpha}n^{\beta}$ as shown in Appendix~\ref{app:A}. Upon integrating over the dimensionless thickness coordinate $\omega^{3}$, the zeroth order term of $\mathcal{E}_{\gamma\,-\,n}$ scales linearly in thickness as follows
\begin{equation}\label{eq:coupling0Result}
    \mathcal{E}^{(0)}_{\gamma\,-\,n} = t\,\gamma^{(0)}_{\alpha\beta}\big[g_{(0)}^{\alpha\beta} + \mathrm{sinc}(2\pi\beta)\big(\cos2\psi\,\Delta_{(0)}^{\alpha\beta} + \sin2\psi\,S_{(0)}^{\alpha\beta}\big)\big],
\end{equation}
where we used trigonometric angle addition formulas and exploited even/odd parity arguments.

The linear order in $\omega^{3}$ receives contributions from two sources, as explicitly shown in Appendix~\ref{app:A}, which eventually gives rise to
\begin{align}\label{eq:coupling1Result}
\mathcal{E}^{(1)}_{\gamma\text{-}n} = t^{2}\,f(\beta)\Big[
    &\,\gamma^{(0)}_{\alpha\beta}\big(\cos2\psi\,S_{(1)}^{\alpha\beta} - \sin2\psi\,\Delta_{(1)}^{\alpha\beta}\big) \\ \nonumber
    &- h_{\alpha\beta}\big(\cos2\psi\,S_{(0)}^{\alpha\beta} - \sin2\psi\,\Delta_{(0)}^{\alpha\beta}\big)
\Big],
\end{align}
where $f(\beta) = ({\sin 2\pi\beta - 2\pi\beta\cos 2\pi\beta})/{2(\pi\beta)^{2}}$. 
Therefore, $\mathcal{E}_{\gamma\,-\,n}$ has the form
\begin{equation}
    \mathcal{E}_{\gamma\,-\,n}=t\,\mathcal{E}^{(0)}\,+\,t^2\,\mathcal{E}^{(1)}+\mathcal{O}(t^3).
\end{equation}
The first term, which is proportional to $t$ represents the in-plane strain contribution to the energy and is already well-understood. Terms of order $t^3$ represent the bending energy. However, we also obtain a contribution to the energy that scales as $t^2$. This anomalous scaling is absent in other KL plate theories and appears purely because of the director's through-thickness rotation.

We can interpret the anomalous term by considering a flat slab with a twisted director through its thickness, as studied by Sawa \textit{et al.} \cite{twistedSelinger}. There, the geometry shows a transition between helicoids, which admit Gaussian curvature, and cylindrical spirals, which do not.
This transition is explained by a similar dimensional reduction that yields a contribution –to the effective $1D$ energy density– linear in the off-diagonal components of the mid-surface curvature $C_{xy}$, with coefficient
$b$ proportional to $\big(sin\,\theta_s\,-\,\theta_s\,cos\,\theta_s\big)/\theta_s^2$ set by the twist angle $\theta_s$ across the thickness. This term, $b\;C_{xy}$, induces the film to twist and the selection between helicoids and spiral ribbons is then governed by a width-to-thickness ratio. 

\begin{figure}[t]
    \centering
    \includegraphics{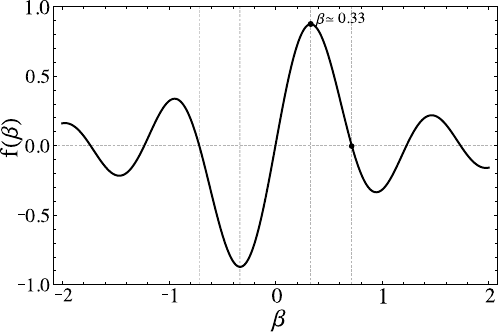}
     \caption{The coefficient $f(\beta)$ of the anomalous $t^2$ as a function of the twist rate $\beta$. The coefficient vanishes as $\beta \rightarrow 0 $ and reaches its first maximum at $\beta \approx 0.33$. The function oscillates and decays with the rapid rotation of the nematic. }
     \label{fig:strengthCoupling}
 \end{figure}

Our $t^2$ contribution –derived from a covariant reduction of the coupling term– plays the same role. Indeed, in Eq. (\ref{eq:coupling1Result}) the curvature, 
couples to the shear $S^{\alpha \beta}$, producing precisely these off-diagonal curvature terms with a coefficient $f(\beta)$ proportional  up to a factor with $b$ of Sawa \textit{et al}. The midsurface is therefore driven toward a twisted configuration.
The magnitude of $f(\beta)$ is shown in Fig.~\ref{fig:strengthCoupling}. As expected, the anomalous term $t^2$ turns off in the absence of twist $\beta=0$. The odd symmetry of $f(\beta)$ possibly reflects the chirality of the director. As a result the $t^2$ term may reverse with the handedness of the director.
Interestingly, the function $f(\beta) \rightarrow 0$ with increasing $\beta$, meaning that when the director rotates many times across the thickness, the $t^2$ coupling is progressively washed out. 

We also note that the integral $\int_{-1}^{+1}\!\omega^{3}\,\sin(2\pi\beta\omega^{3})\,d\omega^{3}$ which gives rise to $f(\beta)$ –under the assumption of rapid twist–, integrates to $\propto t$ when holding the pitch fixed. As a result, the contribution $t^{2}f(\beta)$ then reduces to order $t^3$, joining the usual bending terms and recovering the standard structure of KL plate theories.

\subsection{Linear stability analysis for infinite cylinder}\label{subsec:LSA}

Liquid crystal elastomers present mechanical instabilities that can be harnessed for functionality, originating from the coupling between elasticity and nematic orientation. This coupling opens two distinct routes to instabilities. In the first route, soft elasticity allows a certain set of deformations through director rotation at low energy cost. When the energy landscape is sufficiently flat, the material then becomes susceptible to buckling, wrinkling, stripe formation and the other instabilities discussed below. In the second, the instability emerges when the locally preferred deformation induced by the nematic is incompatible with the global geometry. The resulting frustration can therefore be relieved through the emergence of unstable modes. It is the second route that we explore here. 

The instabilities observed in nematic  elastomers have long been an area of focus. The most commonly characterized example is the stripe-domain instability where the director reorients itself at a critical shear strain \cite{WarnerTerentjevVerwey1996}, observed experimentally for a monodomain nematic elastomer by Kundler and Finkelmann \cite{Finkelmann1997} and given a rigorous framework by DeSimone and Dolzmann \cite{desimone2002}. Beyond stripes, studies have also observed wrinkling in LCE bilayers \cite{Dias2019, GorielyMihai2021, Plucinsky2017}, inflation of LCE balloons \cite{GiudiciBigginsBalloons2020}, and aneurysm in pressurized cylinders \cite{LeeCylinders2021}. 

A director that twists through the thickness is a particularly rich source of geometric frustration. Prior work on twisted nematic geometries has focused on flat geometries with a through-thickness twisted director \cite{twistedSelinger, gammaAgostiniani} as mentioned in Sec.~\ref{subsec:DimRed}. Here, we are concerned with the resulting shape change of a topologically non-trivial geometry, the cylinder, which is fundamentally incompatible with the local twisting. We perform a stability analysis around an equilibrium state through an eigenvalue analysis of the Hessian matrix. Our control parameter is the perturbative parameter $q_0$ which may not be mapped directly to an experimental parameter. We also note that the cylindrical case is computed by direct dimensional reduction of the 3D energy in cylindrical coordinates, which serves as an independent realization of the framework in Sec. \ref{sec:EffectiveTheory}, yet it generates the same $t^2$ term that we noted above.

We linearize the energy about an equilibrium cylinder of radius $r^*(q_0;t,\psi,\Theta)$ obtained pertubatively in the parameter $q_0$. We write the perturbed midsurface as 
\begin{align}\label{eq:perturbedMid}
    x_m(\theta,z)&=(r^*\cos\theta, r^*\sin\theta,z)\\ \nonumber
    &+\epsilon\left(\rho(\theta,z)\hat{\bf N_{0}}+u_\theta(\theta,z)\partial_\theta \mathbf{{x}}^{m}_{0}+u_z(\theta,z)\partial_z\mathbf{{x}}^{m}_{0}\right)
\end{align}
where $\mathbf{{x}}^{m}_{0}=\left(r^{\ast}\cos\theta,\; r^{\ast}\sin\theta,\; z\right)$ and extend it through thickness by the KL hypothesis, and re-perform the dimensional reduction of Sec. \ref{sec:EffectiveTheory}. The induced metric, strain, director frame, and nematic director are expanded to $\mathcal{O}(\zeta^2)$ in the thickness coordinate and integrated, yielding a two-dimensional energy. Expanding the energy in the perturbation amplitude $\mathcal{O}(\epsilon^2)$, the linear contribution vanishes and the quadratic term defines the Hessian. 

In Eq. \ref{eq:perturbedMid}, $\rho, u_\theta, u_z$ are the displacement fields along the radial, azimuthal and longitudinal direction of the cylinder, respectively. Each scalar field is Fourier-decomposed as 
\begin{equation}
    f(\theta,z)=\sum_{m \in \mathbb{Z}}\int dk\; \hat{f}[m, k]\,e^{i(m\theta + kz)},\qquad f \in \{\rho, u_\theta, u_z\}.
\end{equation}
and the Fourier amplitudes construct a $6\times 6$ Hermitian Hessian. The cylinder is linearly stable when the eigenvalues of that matrix are positive; an instability is signaled by the first lowest non-trivial eigenvalue $\lambda_{min}(m,k;q_0)$ crossing zero as $q_0$ increases from 0. The mode $(m^*,k^*)$ at the critical order parameter $q_c$ identifies the shape into which the cylinder first deforms. We note that our tunable, critical parameter is not equivalent to the physical order parameter.

\section{Results}\label{sec:Results}

\subsection{Equilibrium state} \label{subsec:eqRadius}

\begin{figure}[t]
    \centering
    \includegraphics{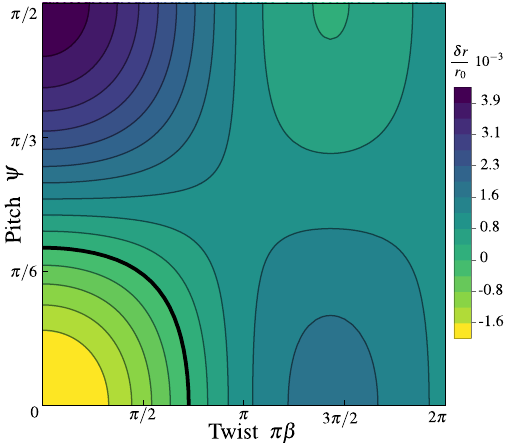}
    \caption{Equilibrium radius shift $\delta r/r_0 = (r^* - r_0)/r_0$ as a function of total twist $\Theta$ and mid-surface pitch $\psi$, at fixed thickness and nematic order ($t = 10^{-3}\, r_0$, $q_0 = 0.1$). The cylinder expands (positive $\delta r$, blue) or contracts (negative $\delta r$, yellow) depending on the combination of $(\Theta, \psi)$. The black curve traces the region where the radial response vanishes.}
    \label{fig:reqContour}
\end{figure}

We will expand around an axisymmetric cylinder whose radius, $r^*$, minimizes the elastic energy. Solving $\partial E/\partial r |_{r = r^*} = 0$ perturbatively to first order in $q_0$ yields the equilibrium radius as a function of physical parameters and the control parameter $q_0$,
\begin{align}\label{eq:eqRadius}
    r^*\paren{q_0;t,\psi,\Theta}
        &= r_0 \nonumber \\
        &+ c_1\, q_0\, \frac{\Theta - 3\cos 2\psi\,\sin\Theta}{\Theta} \\ \nonumber
        &+ c_2\, q_0\, t\, \frac{\sin 2\psi\paren{\sin\Theta - \Theta\cos\Theta}}{\Theta^2},
\end{align}
where $\Theta$, $\psi$ represent  the total rotation of the director across the thickness and the offset respectively, $c_1, c_2$ are set by the Lam\'e constants $\lambda,\mu$. Two structurally distinct contributions appear. The first, of order $t^0$ in the thickness, is generated by the $\mathcal{O}(t)$ term in $E_c$; it oscillates with the pitch as $\cos 2\psi$ and modulates with $\sin\Theta/\Theta$. The second, of order $t$, arises from the $\mathcal{O}(t^2)$ in $E_c$, and vanishes when $\sin 2\psi = 0$.

In the thin-plate limit, the thickness-independent piece dominates and the equilibrium radius is set by the sign of $\Theta - 3\cos 2\psi\,\sin\Theta$. 
Fig.~\ref{fig:reqContour} shows the radius shift $\delta r=r^*-r_0$ as a function of the twist and pitch angles, at fixed thickness and nematic order $q_0$. Blue regions correspond to expansion of the cylinder ($r^* > r_0$) for certain combinations of $(\Theta, \psi)$ and yellow regions to contraction ($r^* < r_0$). The black curve traces the pairs of $\Theta,\psi$, where there is neither contraction nor expansion of the cylinder.

\subsection{First unstable mode}\label{subsec:firstmode}

We examine the eigenspectrum of the Hessian $\mathcal{H}(m,k;q_0,t,\psi,\Theta)$, expanded around $r = r^*$, and identify which mode $(m,k)$ is the first to become unstable with increasing $q_0$. To reduce the number of parameters, we fix the thickness $t = 10^{-3}\, r_0$, midsurface pitch $\psi = \pi/3$, and the twist $\Theta = \pi/6$. Changing these values a little does not affect the qualitative results, but this will be discussed further in Sec.~\ref{subsec:tThetaDep}.

Before assessing stability, we account for the Euclidean motions of the cylinder. With the boundary conditions imposed by the Fourier expansion, these are: the three rigid body translations, and the rigid rotation about the cylinder axis. The other two rotations are not compatible with periodic boundary conditions implicit in the Fourier expansion.
These modes appear at $m=0\,,k=0$ and $m=1\,,k=0$. We constructed these motions explicitly and verified that the apparent negative near-zero eigenvalues are rigid body motions.

\begin{figure}[t]
    \centering
    \includegraphics{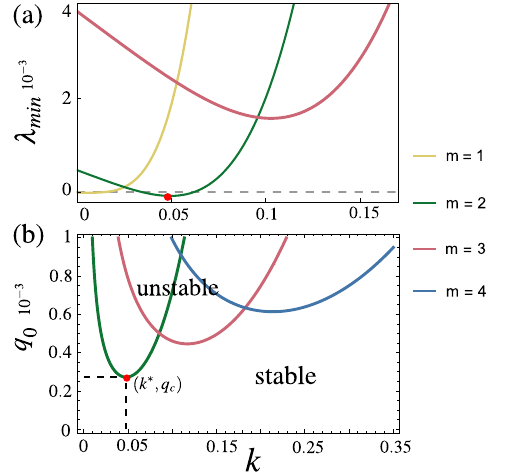}
     \caption{Linear stability of the twisted nematic cylinder at fixed $t = 10^{-3}\, r_0$, $\psi = \pi/3$, $\Theta = \pi/6$. (a)Minimum eigenvalue $\lambda_{\min}(m, k; q_c)$ of the Hessian at the critical nematic order $q_c$, plotted versus axial wavenumber $k$ for $m = 1, 2, 3$. The $m=1$ (dashdotted) branch touches zero at $k=0$ (rigid translation) and rises sharply; the $m=3$ (dashed) branch is positive throughout; the $m=2$ branch (green, solid) curve touches zero at a small finite $k^* \approx 0.05/r_0$ (red dot), identifying the first unstable mode. (b) Stability boundary $\lambda_{\min}(m=2, k; q_0) = 0$ in the $(k, q_0)$ plane. The cylinder is unstable inside the indicated region. The bifurcation point $(k^*, q_c)$ (red dot) marks the onset.}
     \label{fig:DispersionPhase}
 \end{figure}

Fig.~\ref{fig:DispersionPhase}(a) shows $\lambda_{\min}(m,k)$ as a function of the axial wavenumber, $k$, for the three lowest azimuthal wavenumbers $m = 1, 2, 3$, evaluated at the critical value $q_c$ where the spectrum first touches zero. The $m=1$ branch dips to zero at $k=0$ and rises sharply for any $k \neq 0$. 
The $m=2$ and $k\approx0.05$, branch appears to be the first negative eigenvalue (touches zero at $q_c$); this is the first unstable mode. In contrast, the $m=3$ branch has a minimum near $k \approx 0.1$ that remains clearly positive.

Fig.\ref{fig:DispersionPhase}(b) maps the stability boundary $\lambda_{\min}(m^*, k; q_0) = 0$ in the $(k, q_0)$ plane. The cusp at $(k^*, q_c)$ is the bifurcation point of the $m=2$ branch. The fact that $k^* > 0$ rather than $k=0$ is the central qualitative result of the analysis, meaning that the most unstable mode is not the pure ovalization but a long wavelength helical ovalization. In other words, the unstable mode carries two distinct features, an elliptical cross-section and a modulation along the axis, hence the nonzero axial wavenumber. We can also see from Fig.\ref{fig:DispersionPhase}(b) that it goes unstable at $q_c\approx3 \cdot10^{-4}$.
As the twisted director enters the energy through $\mathcal{E}_{\gamma-n}$, it sets a nonzero axial wavenumber, corresponding to $\lambda \approx 125\, r_0$. This reflects a competition between the stretching energy and the anomalous $t^2$ term.


 \begin{figure}[t]
    \centering
    \includegraphics[width=\columnwidth]{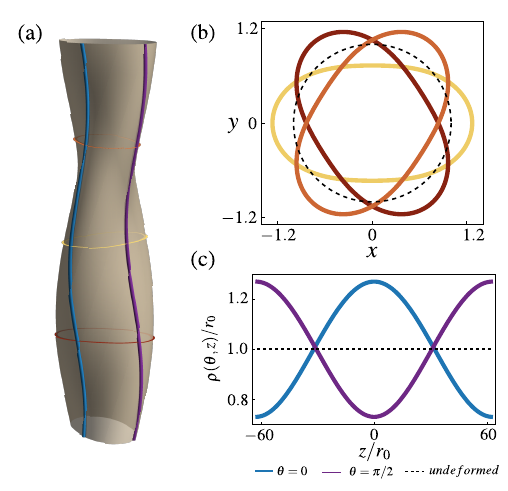}
    \caption{First unstable mode $(m^*, k^*) = (2, 0.05/r_0)$ of the twisted nematic cylinder at $q_0 = q_c$ for $t = 10^{-3}\, r_0$, $\psi = \pi/3$, $\Theta = \pi/6$ (a) Three-dimensional view of the deformed mid-surface for the first unstable mode with three $z$-slices and two waves being traced at $\theta=0,\pi/2$. (b) Cross section of the deformed cylinder for varying z-slices and the undeformed (dashed) circular cross section. The cross sections show the elliptical nature of the $m=2$ wavenumber and the translational symmetry break of the instability along the cylinder axis. (c) Radial distance $\rho(\theta,z)/r_0$ from the cylinder axis at fixed azimuthal angles $\theta=0$ (blue with circle markers), $\theta=\pi /2$ (purple, dashdotted) versus z. The two curves are in antiphase, confirming the $m=2$ ellipticity; the axial oscillation reveals the $k^*$-modulation with wavelength $\lambda = 2\pi/k^* \approx 125\, r_0$.}
    \label{fig:firstUnstableMode}
\end{figure}

The cross-sectional ellipticity and the long-wavelength axial modulation are independent features of the unstable mode, and panels (a)–(c) of Fig.~\ref{fig:firstUnstableMode} display them separately. The eigenvector of the marginal mode at $(m^*, k^*)$ is dominated by the radial Fourier amplitude $\rho_{m,-k} \approx 0.894$, with a substantial tangential component $u_{\theta,m,-k} \approx 0.447$ and a negligible axial component. The dominant radial perturbation takes the form $\rho(\theta,z)=\sum_{p=+-}\rho_{m,k} e^{i(m\theta+pkz)}$, but after imposing the condition $\rho=\rho^*$ for a real displacement field it becomes $\rho(\theta,z)=\rho_{m,-k}~cos(m\theta-kz)$ or $\rho(\theta, z) \propto \cos(m\theta - kz)$. The antiphase relation $\rho(0, z) = -\rho(\pi/2, z)$ at fixed $z$ is the ellipticity of $m=2$, shown directly in the $z$-slices of panel (b): the cylinder is fatter along $\theta = 0$ whereas it is thinner along $\theta = \pi/2$, with the long axis of the ellipse rotating as $z$ advances. The radial displacement at fixed $\theta$ in panel (c) oscillates between $0.75\, r_0$ and $1.25\, r_0$ along the axis with wavelength $2\pi/k^*$. The substantial tangential component $u_\theta$ is what makes the flattening helical rather than a simple translation-invariant ovalization: it couples in-plane shear to the radial deformation in a way that follows the chirality of the underlying twisted director. Panel (a) shows the resulting three-dimensional shape of the midsurface where the elliptical cross-section sweeps along the cylinder axis with the axial modulation set by $k^*$. 

For reference, the $m=2$ ovalization is reminiscent of the classical Brazier effect \cite{Brazier1927}, in 
which a thin-walled cylinder under bending develops an elliptical cross-section. 
The difference is the mechanism: the ovalization is driven by a nematic coupling rather than an external load. 
The helical nature of the deformation, \textit{i.e.} the axial modulation, is also reminiscent of the Helfrich-Hurault instability 
\cite{Helfrich2023}, where a cholesteric liquid crystal develops a periodic undulation along the axis \cite{ChrisPincus}. Both features
–the elliptical cross-section and the axial modulation– are driven by the nematic coupling, and specifically 
by the twisted nematic director that is incompatible with the cylindrical geometry. The instability is therefore a
manifestation of the geometric frustration that arises from the competition between the nematic coupling and the cylindrical geometry.

\subsection{Thickness and twist angle dependence}\label{subsec:tThetaDep}

 \begin{figure}[t]
    \centering
    \includegraphics{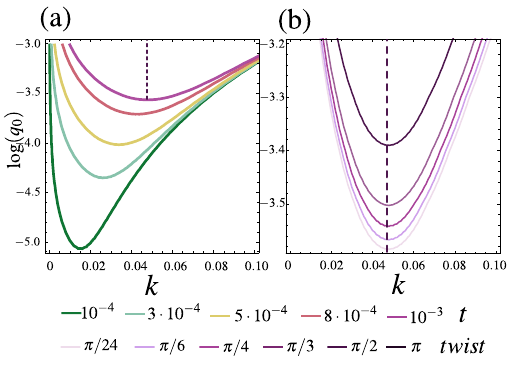}
    \caption{Stability boundary. (a) $\lambda_{\min}(m=2, k; q_0) = 0$ in the $(k, \log q_0)$ plane for five thicknesses $t \in \{10^{-4}, 3\times 10^{-4}, 5\times 10^{-4}, 8\times 10^{-4}, 10^{-3}\}\, r_0$ at fixed $\psi = \pi/3$ and $\Theta = \pi/6$. Each curve is the locus where the $m=2$ branch first crosses zero. Thinner plates become unstable at smaller $q_0$ and at smaller wavenumber $k^*$. (b)$\lambda_{\min}(m=2, k; q_0) = 0$ in the $(k, \log q_0)$ plane for six twist angles $\Theta \in \{\pi/24, \pi/6, \pi/4, \pi/3, \pi/2, \pi\}$ at fixed $\psi = \pi/3$ and $t = 10^{-3}\, r_0$. The marginal wavenumber $k^*$ and azimuthal index $m^* = 2$ are essentially independent of $\Theta$, while the critical order $q_c$ shifts modestly. At $\Theta = \pi$ the curve cannot be seen as Q is double headed.}
    \label{fig:criticalQthickness}
\end{figure}

The bifurcation threshold, $q_c$, and the wavenumber, $k^*$, of the marginal mode depend on two experimentally tunable parameters: the plate thickness $t$ and the total twist angle $\Theta$. Figure~\ref{fig:criticalQthickness} shows the locus of points $\lambda_{\min}(m=2, k; q_0) = 0$ in the $(k, \log q_0)$ plane for a family of thicknesses $t \in \{10^{-4}, 3\times 10^{-4}, 5\times 10^{-4}, 8\times 10^{-4}, 10^{-3}\}\, r_0$, with $\psi$ and $\Theta$ held fixed. Two trends are visible. First, the bifurcation threshold increases monotonically with $t$: thinner plates become unstable at smaller $q_0$. Second, the wavenumber $k^*$ at which the threshold is minimized is smaller at smaller $t$, indicating that thinner cylinders prefer longer-wavelength helical patterns. Across the full range tested, the marginal mode remains $m=2$.
We note that the linear analysis becomes increasingly inadequate away from $q_c$ as it locates the bifurcation but cannot follow the system past the onset, where nonlinear terms and finite-amplitude effects select the eventual equilibrium shape.

Figure~\ref{fig:criticalQthickness} also shows the stability boundary as $\Theta$ varies over $\{\pi/24, \pi/6, \pi/4, \pi/3, \pi/2, \pi\}$ at fixed $t$ and $\psi$. The marginal mode is robust: $(2, k^*)$ remains essentially unchanged across this range, while $q_c$ shifts modestly meaning that the control parameter is only weakly dependent on the twist angle. The $\Theta = \pi$ case is exceptional: a full $\pi$ rotation across the thickness cancels out with the Q-tensor.

\section{Conclusion}\label{sec:Conclusion}

We have derived an effective non-Euclidean plate theory for a thin nematic elastomer whose director twists rapidly through the thickness, by dimensional reduction of a three-dimensional energy. In addition to the standard stretching ($t$) and bending ($t^3$) energies of Kirchhoff--Love plate theory, the reduction produces an anomalous $t^2$ contribution that couples mid-surface curvature to the through-thickness rotation of the director frame. The term is intermediate in thickness scaling between stretching and bending and dominates bending in the thin-plate limit. Its existence relies on the rapid through-thickness rotation, and it is absent from other Kirchhoff--Love reductions of LCE plates. Applying the same theoretical framework we performed a dimensional reduction to a cylinder of radius $r_0$, we computed the equilibrium radius perturbatively in the nematic order parameter $q_0$ and performed a linear stability analysis of the resulting axisymmetric base state. We found that the cylinder is linearly unstable to a long-wavelength helical flattening mode of azimuthal index $m = 2$ at a small but finite axial wavenumber $k^*$. The eigenvector structure is dominated by the radial displacement with a substantial azimuthal-shear component and it inherits the chirality of the twisted director and produces an elliptical cross-section whose long axis co-rotates with the director along the cylinder axis. The bifurcation threshold $q_c$ is weakly dependent to thickness and twist angle changes.

Beyond these results, our model allows for the study of shape morphing in twisted nematic elastomers and enables the design of tunable surfaces by leveraging instabilities, with potential applications in soft robotics.
This work can be extended in several other directions. The linear analysis locates the bifurcation but cannot follow the system into the nonlinear regime where competition between multiple unstable modes selects the eventual equilibrium shape; direct numerical minimization beyond onset would clarify the post-bifurcation morphology. The long-wavelength character of the marginal mode ($\lambda \approx 125\, r_0$) makes finite-length effects experimentally relevant, and a parallel calculation with explicit end boundary conditions would determine the aspect ratio required to observe the predicted helical flattening in a real sample. Finally, the regime we have studied – where the elastic and coupling energies dominate and Frank elasticity is negligible - can be inverted in samples, where director anchoring and gradient terms enter, opening a different competition relevant to biological systems with helical fiber architectures.

\begin{acknowledgments}\label{sec:Acknowledgments}
We are grateful to Sourav Roy for many discussions regarding this project. We would also like to thank Carlos Enrique Moguel-Lehmer for their helpful comments. This work was supported by Syracuse University and partly through National Science Foundation grant CMMI 2247095.
\end{acknowledgments}

\bibliographystyle{unsrt} 
\bibliography{refs}

\appendix

\section{Expansion of nematic-strain term}\label{app:A}

We decompose the coupling term $\mathcal{E}_{\gamma-Q}^{3D}$
\begin{equation}
   \eta\, \strain{i}{j} Q^{ij} = q\strain{\alpha}{\beta}^{3D}n^\alpha n^\beta ~-~\tfrac{1}{3} q \strain{\alpha}{\beta}^{3D}g^{\alpha \beta}_{3D}~-~\tfrac{1}{3} q \strain{3}{3}^{3D}g^{33}_{3D}.
\end{equation}

 We focus on expanding the strain-nematic term, $\mathcal{E}_{\gamma-n}$, which gives order-by-order the following pieces
\begin{align}
    (\omega^{3})^{0}: \quad &
    \gamma^{3D(0)}_{\alpha\beta}(n^{\alpha}n^{\beta})_{(0)},
    \\[2pt]
    (\omega^{3})^{1}: \quad &
    \gamma^{3D(0)}_{\alpha\beta}(n^{\alpha}n^{\beta})_{(1)}
    - h_{\alpha\beta}(n^{\alpha}n^{\beta})_{(0)},
    \\[2pt]
    (\omega^{3})^{2}: \quad &
    \underbrace{\gamma^{3D(1)}_{\alpha\beta}}_{\propto\, t\, h_{\alpha\beta}}
    (n^{\alpha}n^{\beta})_{(1)}
    \\
    &\quad
    + \gamma^{3D(2)}_{\alpha\beta}(n^{\alpha}n^{\beta})_{(0)}
    + \gamma^{3D(0)}_{\alpha\beta}(n^{\alpha}n^{\beta})_{(2)}
\end{align}
where we ignore second order terms because they give rise to higher $t$ contributions that compete with bending. 

We now integrate over the physical thickness coordinate $\zeta=\omega^{3} t$ with $d\zeta=td\omega^3$ and $\omega^3\in[-1,1]$. The integration of the 
zeroth order term reads as follows
\begin{equation}
    \mathcal{E}_{\gamma - n}^{(0)} = t \int_{-1}^{+1} \gamma^{3D(0)}_{\alpha\beta}\,(n^{\alpha}n^{\beta})^{(0)} \sqrt{|\bar{g}|}\, d\omega^{3}
    \;\sim\; \mathcal{O}(t),
\end{equation}
which can be written as
\begin{equation}\label{eq:intCoupling0}
    \mathcal{E}^{(0)}_{\gamma\,-\,n} = t\,\tfrac{1}{2}\,\gamma^{(0)}_{\alpha\beta}\!\int_{-1}^{1}\!\!\Big(g_{(0)}^{\alpha\beta} + \cos2\Phi\,\Delta_{(0)}^{\alpha\beta} + \sin2\Phi\,S_{(0)}^{\alpha\beta}\Big)\,d\omega^{3},
\end{equation}
where $\gamma^{(0)}_{\alpha\beta}$ and the frame tensors are independent of the thickness coordinate, the only $\omega^3$ dependence resides in the twist angle $\Phi$. We therefore exploit trigonometric arguments using the sum formulas as well as parity arguments such that
\begin{equation}
    \int_{-1}^{+1}\cos 2\varphi\, d\omega^{3} \neq 0,
    \qquad
    \int_{-1}^{+1}\sin 2\varphi\, d\omega^{3} = 0.
\end{equation}
and eventually arrive to Eq.~\ref{eq:coupling0Result}

As for the linear order in $\omega^{3}$ term, there are two contributions as seen above. To solve the integrals we used the parity of  $\w{3}\cos(\pi\beta \w{3})$ (even $\times$ odd = odd) and $\w{3}\sin(\pi\beta \w{3})$ (odd $\times$ odd = even) and eventually arrived to Eq.~\ref{eq:coupling1Result}.

\end{document}